\documentclass[preprint]{aastex61}
\pdfoutput=1

\usepackage{amsmath}
\usepackage{tabularx}
\usepackage{txfonts}
\usepackage{natbib}
\bibpunct{(}{)}{;}{a}{}{,}

\def \gray {gamma-ray }

\def \agile {\it AGILE}
\def \bfn { }
\def \textcl {}

\shorttitle{AGILE observation of ICECUBE-160731}
\shortauthors{Lucarelli et al.}

\begin{document}

\title{{\agile} detection of a candidate gamma-ray precursor to the 
ICECUBE-160731 neutrino event}

\correspondingauthor{Fabrizio Lucarelli}
\email{fabrizio.lucarelli@asdc.asi.it}

\author{F. Lucarelli}
\affiliation{ASI Science Data Center (ASDC), Via del Politecnico snc, I-00133 Roma, Italy}
\affiliation{INAF--OAR, via Frascati 33, I-00078 Monte Porzio Catone (Roma), Italy}

\author{C. Pittori}
\affiliation{ASI Science Data Center (ASDC), Via del Politecnico snc, I-00133 Roma, Italy}
\affiliation{INAF--OAR, via Frascati 33, I-00078 Monte Porzio Catone (Roma), Italy}

\author{F. Verrecchia}
\affiliation{ASI Science Data Center (ASDC), Via del Politecnico snc, I-00133 Roma, Italy}
\affiliation{INAF--OAR, via Frascati 33, I-00078 Monte Porzio Catone (Roma), Italy}

\author{I. Donnarumma}
\affiliation{Agenzia Spaziale Italiana (ASI), Via del Politecnico snc, I-00133 Roma, Italy}

\author{M. Tavani}
\affiliation{INAF/IAPS--Roma, Via del Fosso del Cavaliere 100, I-00133 Roma, Italy}
\affiliation{Univ. ``Tor Vergata", Via della Ricerca Scientifica 1, I-00133 Roma, Italy}
\affiliation{Gran Sasso Science Institute, viale Francesco Crispi 7, I-67100 L`Aquila, Italy}

\author{A. Bulgarelli}
\affiliation{INAF/IASF--Bologna, Via Gobetti 101, I-40129 Bologna, Italy}

\author{A. Giuliani}
\affiliation{INAF/IASF--Milano, via E.Bassini 15, I-20133 Milano, Italy}

\author{L. A. Antonelli}
\affiliation{ASI Science Data Center (ASDC), Via del Politecnico snc, I-00133 Roma, Italy}
\affiliation{INAF--OAR, via Frascati 33, I-00078 Monte Porzio Catone (Roma), Italy}

\author{P. Caraveo}
\affiliation{INAF/IASF--Milano, via E.Bassini 15, I-20133 Milano, Italy}

\author{P. W. Cattaneo}
\affiliation{INFN--Pavia, Via Bassi 6, I-27100 Pavia, Italy}

\author{S. Colafrancesco}
\affiliation{University of Witwatersrand, Johannesburg, South Africa}
\affiliation{INAF--OAR, via Frascati 33, I-00078 Monte Porzio Catone (Roma), Italy}

\author{F. Longo}
\affiliation{Dip. di Fisica, Universita’ di Trieste and INFN, Via Valerio 2, I-34127 Trieste, Italy}

\author{S. Mereghetti}
\affiliation{INAF/IASF--Milano, via E.Bassini 15, I-20133 Milano, Italy}

\author{A. Morselli}
\affiliation{INFN--Roma Tor Vergata, via della Ricerca Scientifica 1, 00133 Roma, Italy}

\author{L. Pacciani}
\affiliation{INAF/IAPS--Roma, Via del Fosso del Cavaliere 100, I-00133 Roma, Italy}

\author{G. Piano}
\affiliation{INAF/IAPS--Roma, Via del Fosso del Cavaliere 100, I-00133 Roma, Italy}

\author{A. Pellizzoni}
\affiliation{INAF -- Osservatorio Astronomico di Cagliari, via della Scienza 5, I-09047 Selargius (CA), Italy}

\author{M. Pilia}
\affiliation{INAF -- Osservatorio Astronomico di Cagliari, via della Scienza 5, I-09047 Selargius (CA), Italy}

\author{A. Rappoldi}
\affiliation{INFN--Pavia, Via Bassi 6, I-27100 Pavia, Italy}

\author{A. Trois}
\affiliation{INAF -- Osservatorio Astronomico di Cagliari, via della Scienza 5, I-09047 Selargius (CA), Italy}

\author{S. Vercellone}
\affiliation{INAF Oss. Astron. di Brera, Via E. Bianchi 46, I-23807 Merate (LC), Italy}

\begin{abstract}
On July 31st, 2016, the ICECUBE collaboration reported the detection
of a high-energy starting event induced by an astrophysical neutrino. 
We report here about the search for a gamma-ray counterpart of the 
ICECUBE-160731 event made with the {\agile} satellite. 
No detection was found {\bfn spanning} the time interval of $\pm 1$~ks around the 
neutrino event time $T_0$ {\bfn using the {\agile} ``burst search'' system.}
{\bfn Looking for a possible gamma-ray precursor in the results of 
the {\it AGILE}-GRID automatic {\it Quick Look} procedure over predefined 
48-hours time-bins,} we found an excess above 100 MeV between 
one and two days before $T_0$, positionally consistent with the 
ICECUBE error circle, having \textcl{\bfn a post-trial} significance 
of about $4\sigma$.
\textcl{\bfn A refined data analysis of this excess confirms {\it a-posteriori} the 
automatic detection. The new {\agile} transient source, named 
AGL J1418+0008, thus stands as possible ICECUBE-160731 gamma-ray precursor.}
{\bfn No other space missions nor ground observatories have reported any 
detection of transient emission consistent with the ICECUBE event.
We show that Fermi-LAT had a low exposure of the ICECUBE region 
during the {\agile} \gray transient.}
Based on an extensive search for cataloged sources within 
the error regions of ICECUBE-160731 and AGL J1418+0008, we find 
a possible common counterpart showing some of the key features associated 
to the high-energy peaked BL Lac (HBL) class of blazars. Further 
investigations on the nature of this source using dedicated 
SWIFT ToO data are presented.
\end{abstract}


\section{Introduction}
Neutrino astronomy by under-water and under-ice Cherenkov detectors has 
entered a new era since the completion of the ICECUBE and ANTARES 
telescopes~\citep{2010RScI...81h1101H, 2011NIMPA.656...11A} 
and the subsequent first clear detection of a diffuse background 
of Very High Energy (VHE) extra-terrestrial 
neutrinos~\citep{2013Sci...342E...1I, 2015PhRvL.115h1102A}. 
{\bfn No significant clustering of neutrinos above background expectation has been 
observed yet~\citep{2017ApJ...835..151A}, although the ICECUBE 
apparatus might reach the sensitivity 
or accumulate enough statistics to unambiguously detect anisotropy 
or clustering of events within a few more years of observations.}

Emission of TeV-PeV neutrinos might be due to exceptionally energetic 
transient phenomena like flaring activities from Active Galactic Nuclei (AGNs), 
Gamma-Ray Bursts (GRBs) or Supernovae explosions~\citep{Anchordoqui:2013dnh}. 
A direct correlation between gamma-rays and neutrinos from astrophysical 
sources is expected whenever hadronic emission mechanisms are at work. 
In particular, several theoretical works assume that neutrinos production 
occurs in astrophysical beam dumps, where cosmic rays accelerated in regions 
of high magnetic fields near black holes or neutron stars interact via proton-proton
({\it pp}) or proton-photon ({\it p$\gamma$}) collisions with the matter or the radiation 
field surrounding the central engine or in a jet of plasma ejected from it, 
giving raise also to gamma-rays emission (see \citep{2017NatPh..13..232H} 
for a review).

Supernovae remnants (SNRs) expanding in dense molecular clouds and 
microquasars in our Galaxy as well as AGNs of the blazars category 
are the main neutrino source candidates up to PeV 
energies~\citep{1989A&A...221..211M, 1995APh.....3..295M, 
1997ApJ...488..669H, 1998APh.....9....1P, 2005ApJ...631..466B, 
2006APh....26..310V, 2014ApJ...780...29S}. 
Besides the identification of the {\it pion excess} in \gray 
observations of SNRs interacting with molecular 
clouds~\citep{2011ApJ...742L..30G, 2013Sci...339..807A}, 
detection and identification of a clear neutrino point-like 
source would represent the evidence of proton and hadron 
acceleration processes, resolving as well the long-lasting problem of the 
cosmic ray origin (at least up to multi-PeV energies).

{\bfn Since April 2016}, the ICECUBE experiment alerts almost 
in {\it real time} the astronomical community whenever an extremely 
high-energy single-track neutrino event (with energy in the sub-PeV to PeV range) is 
recorded. The communication is sent through the ICECUBE\_HESE 
(a single high-energy starting ICECUBE neutrino) and the ICECUBE\_EHE 
(extremely high-energy ICECUBE neutrino) GCN/AMON notices 
system~\citep{KEIVANI2016} a few seconds after the event 
trigger. The instant notice provides a first determination 
of the statistical relevance of the event and the 
reconstructed neutrino arrival direction, projected onto the sky, with 
its 90\% and 50\% {\bfn containment radius (c.r)}\footnote{\bfn For ICECUBE\_EHE notices, 
only source errors at 50\% c.r. are given.}.

On July 31st, 2016, the ICECUBE {\bfn Collaboration reported} a HESE GCN/AMON 
notice\footnote{http://gcn.gsfc.nasa.gov/notices\_amon/6888376\_128290.amon} 
{\bfn announcing} the detection of a high-energy neutrino-induced 
track-like event at time $T_0$ = 01:55:04.00 UT (MJD=57600.07990741).
The event was also classified as EHE event, {\bfn possibly having an 
energy higher than several hundred TeV\footnote{\bfn 
As quoted in the ICECUBE\_EHE event information web page 
https://gcn.gsfc.nasa.gov/amon\_ehe\_events.html}} and a {\it signalness}\footnote{
Probability that the neutrino event is of astrophysical origin.} 
of $\sim$~0.85. This neutrino detection triggered a broad-band 
follow-up by several space and ground-based instruments, searching 
for an electromagnetic (e.m.) counterpart to be associated to the neutrino emission.

In what follows, we report about the search for a \gray counterpart of 
the ICECUBE-160731 neutrino event made using the data of the 
{\it AGILE} satellite. The paper is organized as follows: in Section 2, we 
describe the main {\it AGILE} instrumental characteristics and its unique 
capabilities for the search of gamma-ray counterparts to such triggered events 
of very short duration. In Section 3, we present the results 
of the {\it AGILE} observations, both near the prompt neutrino 
event time $T_0$ and in archival data. In Sections 4, we report about the 
multi-wavelength (MWL) follow-up and in Section 5 we search for a possible 
e.m. counterpart candidate using the cross-catalog search tools available 
from the ASI Science Data Center (ASDC)\footnote{http://www.asdc.asi.it}. 

\section{{\agile} as detector of transient gamma-ray sources}
The \gray satellite {\it AGILE}~\citep{2009A&A...502..995T}, 
launched on 2007, has just completed its tenth year of operations in orbit. 
The main on-board instrument is the \gray imaging detector (GRID) 
sensitive to gamma-rays in the energy range 30 MeV--50 GeV, composed 
by the \gray Silicon Tracker, the Mini-Calorimeter (MCAL) and the
anti-coincidence (AC) system for the particle background rejection.
The co-axial X-ray (20-60 KeV) detector Super-{\agile} completes 
the satellite scientific payload.

Since Nov. 2009, {\it AGILE} is operated in the so called 
{\it spinning} observation mode, in which the satellite rotates 
around the Sun-satellite versor. In this operation mode, the {\it AGILE} \gray 
imager approximately observes the whole sky every day, with a 
sensitivity (at 5$\sigma$ detection level) to \gray fluxes above 100~MeV 
of the order of $(3 \div 4) \times 10^{-6} \rm~ph~cm^{-2}~s^{-1}$. 

As already demonstrated in the recent follow-up of the gravitational-wave 
event GW150914~\citep{2016ApJ...825L...4T} and in dozens 
of Astronomer's Telegrams (ATel) and GCN circulars, 
{\it AGILE} is a very suitable instrument to perform searches 
for short transient \gray sources and \gray counterparts to 
multi-messenger transient events like the neutrino event 
observed on July 31st, 2016.

The main characteristics that make {\agile} in spinning mode an important
instrument for follow-up observations of multi-messenger counterparts are:
\begin{itemize}
\setlength{\itemsep}{0pt}
 \item a very large {\bfn field of view (FoV)} of 2.5~sr for the {\agile}-GRID;
 \item best sensitivity to \gray fluxes above 30 MeV of the order 
of $(2 \div 3) \times 10^{-4} \rm~ph~cm^{-2}~s^{-1}$ for typical single-pass 
integrations of 100~s;
\item a coverage of 80\% of the whole sky every 7 minutes;
 \item a gamma-ray exposure of $\sim$2 minutes of any field in the 
accessible sky every 7 minutes;
\item between 150-200 passes every day for any region in the accessible sky. 
\item sub-milliseconds trigger for very fast events.
\end{itemize}

Despite the small size {\bfn (approximately a cube of side $\sim$ 60 cm)}, 
the {\agile}-GRID achieves an effective area of the order of 500~$\rm cm^2$ between 
200~MeV and 10~GeV for on-axis gamma-rays, and an angular resolution 
(FWHM) of the order of 4$^\circ$ at 100~MeV, decreasing 
below 1$^\circ$ above 1~GeV~\citep{2011NIMPA.630..251C, 2013A&A...558A..37C, 
2015ApJ...809...60S}.

A very fast ground segment alert system allows the AGILE Team 
to perform the full {\agile}-GRID data reduction and the preliminary 
{\it Quick Look (QL)} scientific analysis only 25/30 minutes 
after the telemetry download from the spacecraft~\citep{2013NuPhS.239..104P, 
2014ApJ...781...19B}.

{\bfn The {\it AGILE QL} on-ground system implements two different 
kinds of automatic analysis:
\begin{itemize}
\item A ``burst search'' system, involving both GRID and MCAL instruments, 
is used to look for transients and GRB-like phenomena 
on timescales ranging from a few seconds to tens of seconds\footnote{\bfn A 
special sub-millisecond search for transient events detected by MCAL is
operational on board~\citep{2009A&A...502..995T}.}. The burst search system 
runs on predefined time windows of 100 seconds, and it may be also triggered 
by external GCN notices~\citep{ZOLI2016}.
\item A ``standard'' {\agile}-GRID {\it QL} analysis, based on a Maximum Likelihood 
(ML) algorithm~\citep{1996ApJ...461..396M, 2012A&A...540A..79B}, is used to detect 
gamma-ray transients above 100 MeV on timescales of 1-2 days~\citep{2014ApJ...781...19B}. 
This automatic procedure routinely runs over predefined 48-hours time-bins. 
\end{itemize}

Given the {\agile} effective area and sensitivity, these 
collecting time intervals are the most appropriate to accumulate 
enough statistics and to maximize the signal-to-noise ratio in both cases.}

\section{{\agile} investigations of ICECUBE-160731}
\begin{figure}[t]
  \centering
  \includegraphics[width=12cm]{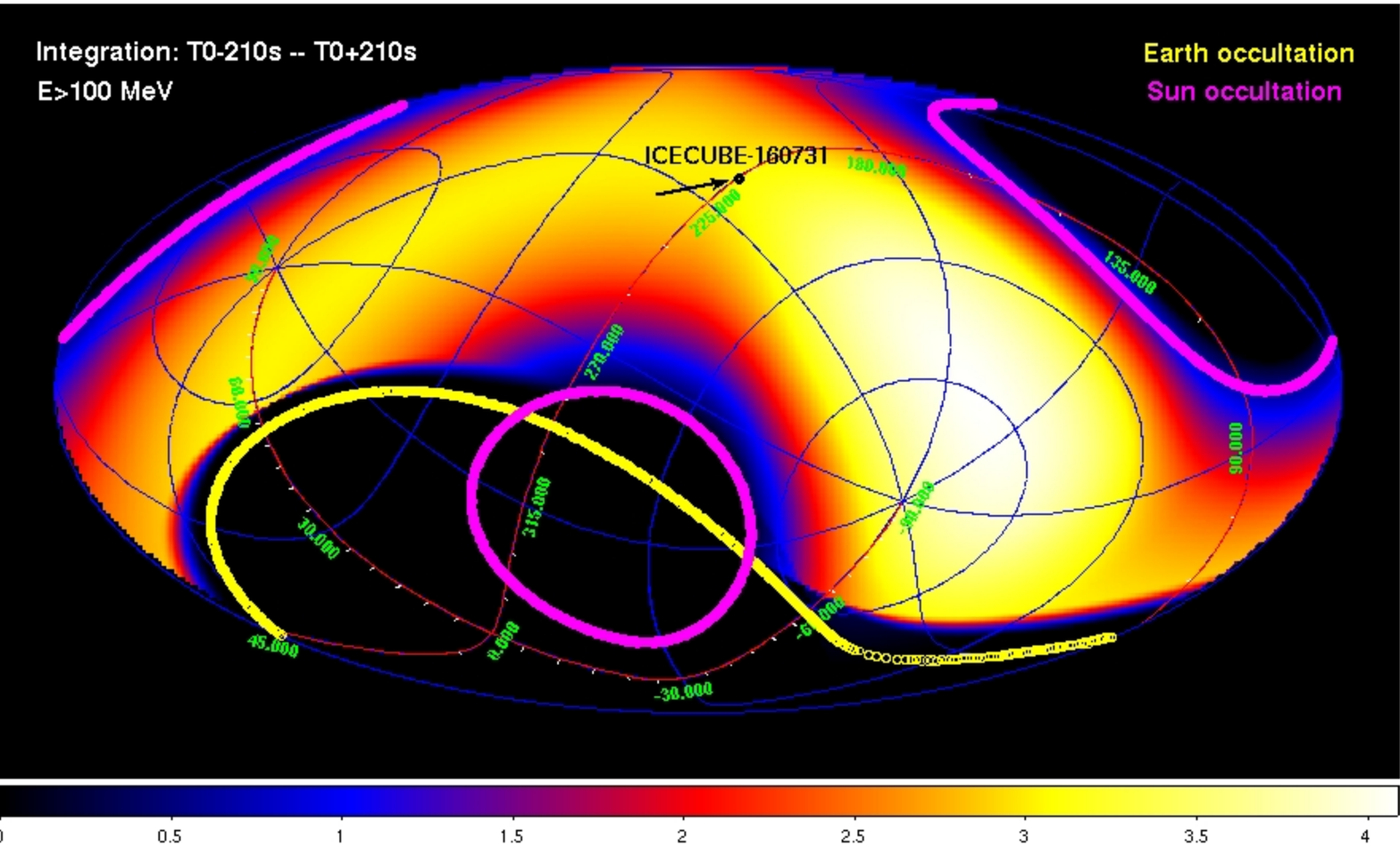}
  \caption{Hammer-Aitoff projection, in Galactic coordinates, of the 
{\agile} \gray exposure in [$\mathrm{cm^2~s~sr}$] (bin size of 0.5$^\circ$) 
after one complete rotation in spinning mode, time-centered at the 
ICECUBE-160731 event time $T_0$. The neutrino event error circle 
is shown in black. The magenta and yellow contours show, respectively, the Sun/anti-Sun 
exclusion regions and the average Earth occultation during 
the considered integration time: ($T_0$-210; $T_0$+210)~s.}
  \label{figure_onerotexp}
\end{figure}

The ICECUBE-160731 best-fit reconstructed neutrino arrival direction in 
equatorial coordinates is (from Rev. \#1 of the GCN notice):
\begin{quote} R.A.,Dec (J2000)=(214.5440, -0.3347) +/- 0.75 [deg] \end{quote}
(90\% statistical plus systematic c.r.), corresponding to 
Galactic coordinates: l,b=(343.68, 55.52) [deg].
{\bfn In the next sections, details of the automatic and refined {\agile} data 
analysis of the ICECUBE-160731 event are reported.}

\subsection{Prompt event} \label{prompt_event}
{\bfn The search for a GRB-like prompt event on short time-scales ranging from 
a few to tens of seconds connected to the ICECUBE neutrino emission was performed
with the {\agile} burst search system. The system was triggered by the first 
ICECUBE GCN/AMON notice reported a few tens of seconds after $T_0$. 
The automatic procedure searches for prompt \gray emission 
on predefined 100~s time-interval bins ranging from 
$T_0$-1000 to $T_0$+1000~s. On these short timescales, the 
method of the ML is not applicable, and an aperture photometry 
is applied. The significance of the signal with respect to the 
background is calculated using the Li\&Ma formula~\citep{1983ApJ...272..317L}.}

Near $T_0$, the reconstructed neutrino-source position was in 
good visibility for the {\agile}-GRID FoV, neither occulted by 
the Earth nor by the exclusion regions around the Sun and anti-Sun 
positions (see Fig.~\ref{figure_onerotexp}). No significant 
detection was found in the GRID data from the event position 
in any of the 100~s time-bins scanned. The 3$\sigma$ Upper Limit (UL) 
for the emission in the range 30 MeV--50 GeV estimated in the 
100~s time-bin with the highest exposure on the event position is: 
$5.7 \times 10^{-4} \mathrm{~ph~cm^{-2}~s^{-1}}$.

Moreover, using the data of the {\agile}-MCAL and the AC scientific 
ratemeters, we have searched for burst-like events in the energy range of 
0.4 -- 100~MeV and 70 keV -- tens of MeV, respectively. No significant 
event has been detected in neither of the two detectors. 

\subsection{Search for \gray precursor and delayed emission}\label{precursor}
Since the astrophysics and the time scales of the phenomena 
related to the emission of these extremely high-energy neutrinos are 
still uncertain, besides the investigations near $T_0$ 
we also explored the {\it AGILE}-GRID data taken few days before and 
after $T_0$, searching for a possible \gray precursor or 
delayed emission on longer (daily) time-scales possibly connected 
to the neutrino event. 

\textcl{\bfn Interestingly, a \gray excess above 100~MeV with a 
\textcl{\bfn pre-trial ML} significance of 4.1$\sigma$ compatible 
with the ICECUBE error circle appeared in the results of 
the {\agile}-GRID automatic {\it QL} procedure between one and 
two days before $T_0$. This detection was reported in 
the ATel \#9295~\citep{2016ATel.9295....1L}.}

\textcl{\bfn The automatic {\it AGILE} QL procedure 
runs on predefined 2-days integration time since Nov. 2009, 
the starting of the spinning observation mode.} 
\textcl{\bfn The AGILE source ML detection method 
derives, for each candidate source, the best parameter estimates 
of source significance, \gray flux, and source location. 
The ML statistical technique, used since the analysis 
of EGRET \gray data~\citep{1996ApJ...461..396M} and adapted 
to the {\agile} data analysis~\citep{2012A&A...540A..79B, AGILESW}, compares 
measured counts in each pixel with the predicted counts derived 
from the diffuse \gray model to find statistically significant 
excesses consistent with the instrument point spread function.}

\textcl{\bfn An {\agile} {\it QL} “detection” is in general 
defined by the condition $\sqrt{TS}\geq4$, where {\it TS} is the Test 
Statistic of the ML method defined as 
$-2 log(\mathcal{L}_0/\mathcal{L}_1)$, where $\mathcal{L}_0/\mathcal{L}_1$ 
is the ratio between the maximum likelihood of the null hypothesis over 
the point-like source hypothesis, given the diffuse {\agile} \gray background 
model~\citep{2004MSAIS...5..135G}. This threshold has been calibrated 
over various timescales and different background conditions 
(e.g., on or outside the Galactic plane)~\citep{2012A&A...540A..79B}.}

{\bfn To evaluate the post-trial significance of the automatic 
QL detection mentioned above}, we used the probability distribution of the ML Test 
Statistic (TS) computed in \cite{2012A&A...540A..79B}. The probability of having 
at least one detection {\bfn due to a background fluctuation 
for any position within the predefined Region of Interest 
(ROI) of 10$^\circ$ radius} used in the ML fitting procedure with a significance 
$\sqrt{TS} \ge h$, in $N$ independent trials, 
\textcl{\bfn is given by $P_1(N)=1-(1-p)^{N}$}, where $p$ is 
the $p$-value {\bfn (that is, the probability of finding a false positive detection 
in a single observation)} corresponding to $h$. The $p$-value for a 
detection with $\sqrt{TS} \ge 4.1$ outside the Galactic plane\footnote{
{\bfn As expected by the Wilks' theorem~\citep{Wilks}}, the TS values 
follow in this case the $\frac{1}{2}\chi^2$ distribution with one 
degree of freedom.} is $3.8\times 10^{-5}$. 
{\bfn By considering all the generated maps having enough exposure in 
spatial coincidence with the neutrino error circle (amounting to 226 
since the beginning of the spinning observation mode), 
the probability of having one detection by chance in $N$=226 trials 
is $P_1(226)=8.5 \times 10^{-3}$. The chance probability of the 
{\agile} detection becomes at least two orders of magnitudes lower 
if we consider the probability $P_2$ of spatial 
coincidence of the {\agile}-GRID excess with the ICECUBE error region 
within the 10$^\circ$ radius ROI. The combined post-trial probability becomes then 
$P_1 \times P_2 \sim 8.5 \times 10^{-5}$, which corresponds to 
a 3.9$\sigma$ post-trial significance.}

\textcl{\bfn A refined analysis has been performed both to confirm the 
automatic {\it QL} result (applying more stringent cuts to further reduce the background 
contamination from albedo events) and to find a better 
temporal characterization of the gamma-ray transient positionally 
consistent with the ICECUBE-160731 position.}

\begin{figure} [t!]
  \centerline{
    \includegraphics[width=12cm]{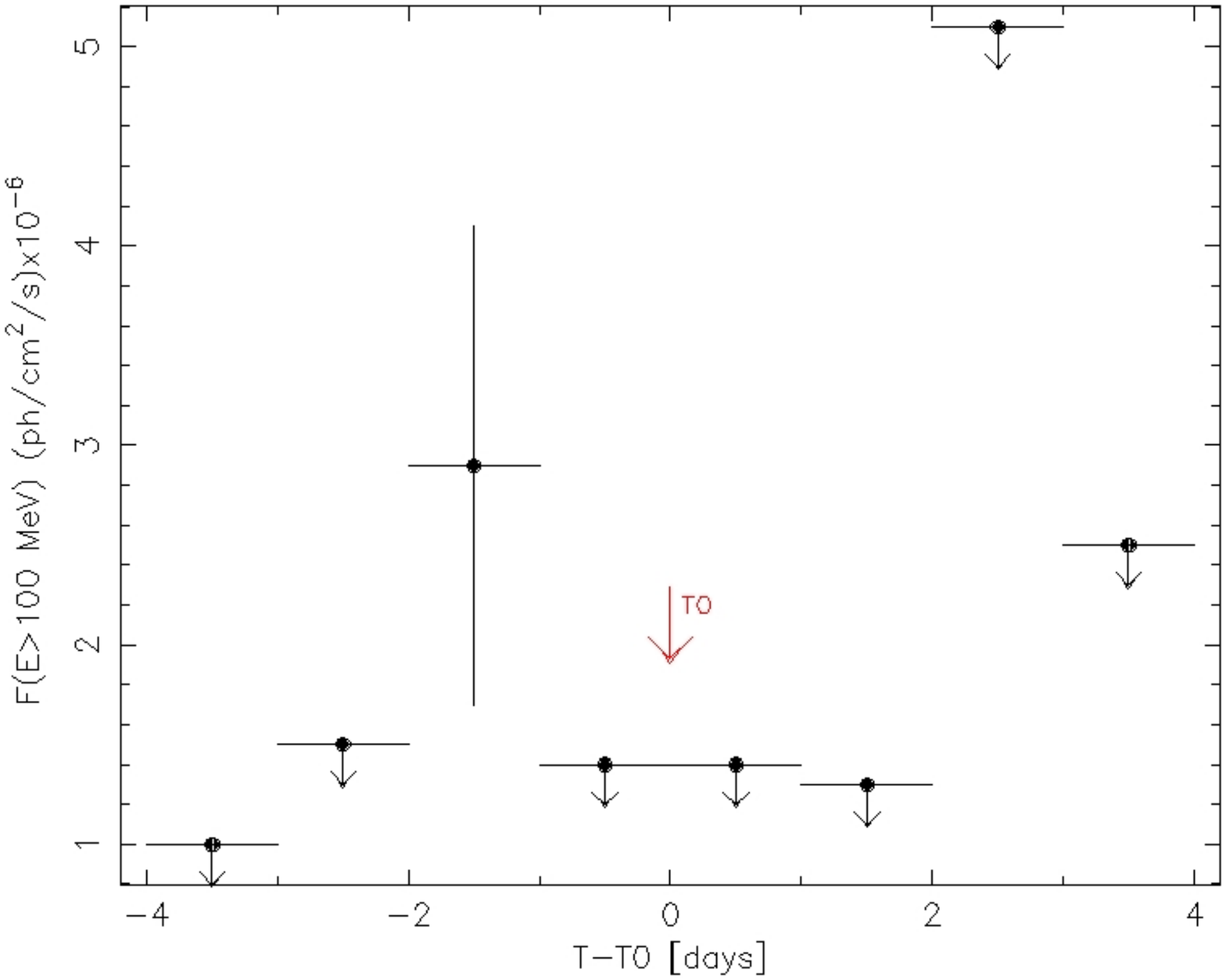}}
  \caption{A-posteriori refined analysis: 
{\agile}-GRID 1-day time-bin lightcurve starting at $T_0$-4 days 
(MJD=57596.07991) obtained from the {\agile} ML analysis performed at the 
ICECUBE-160731 position \textcl{\bfn over each integration bin}.}
\label{grid_lc}
\end{figure}

\textcl{\bfn In the refined GRID data analysis, we created a 
lightcurve symmetric with respect to T0, using 
a time-bin of 24 hours, which is the minimum integration 
time needed by the GRID to detect a 
medium/high flaring gamma-ray source above 100~MeV with enough 
statistics\footnote{\textcl{\bfn Only in some 
exceptional bright flares the integration time-bin may be reduced 
below 24 hours (see, e.g., \cite{2011ApJ...741L...5S, 
2011ApJ...736L..38V})}.}.}

A search for \gray emission above 100~MeV using the {\agile} ML 
around the ICECUBE position has \textcl{\bfn thus} been performed over 
the time interval ($T_0$-4; $T_0$+4)~days. 
Exposure, counts and diffuse emission maps \textcl{\bfn of each time-bin} were generated 
using the official {\agile} scientific analysis software (release: BUILD 21; 
response matrices: I0023)~\footnote{\tt http://agile.asdc.asi.it/public/AGILE\_SW\_5.0\_SourceCode/}~\citep{AGILESW}, applying a cut of 90$^\circ$ on the albedo events rejection parameter 
and taking an {\it AGILE}-GRID FoV radius of 50$^\circ$. {\bfn In comparison, 
the predefined {\it QL} maps are generated with a looser albedo cut of 80$^\circ$ 
and a larger acceptance FoV radius of 60$^\circ$.}
GRID data acquisition during the passage over the South Atlantic Anomaly (SAA) 
is suspended. \textcl{\bfn Each time-bin of the lightcurve has been 
analyzed by means of the ML algorithm assuming a gamma-ray source 
at the ICECUBE position. Figure~\ref{grid_lc} shows the \textcl{\bfn resulting} \gray light-curve, 
where for each bin, the ML gamma-ray flux estimate above 100 MeV or the 95\% C.L. UL at 
the input ICECUBE-160731 position is shown.}

A gamma-ray excess above 100 MeV with a ML significance 
of 4.1$\sigma$ is detected in the bin centered one day and a half 
before the $T_0$ (from MJD=57598.07991 to MJD=57599.07991), confirming the 
automatic {\it QL} detection~\citep{2016ATel.9295....1L}. 
The candidate gamma-ray precursor has an estimated flux of \begin{quote} 
F(E$>$100~MeV)=$(3.0 \pm 1.2) \times 10^{-6} \, \rm ph \, cm^{-2} \, s^{-1}$ \end{quote}
with centroid Galactic coordinates \begin{quote} l,b=(344.01, 56.03) 
$\pm$ 1.0 [deg] (95\% stat. c.l.) $\pm$ 0.1 [deg] (syst.), \end{quote}
compatible with the ICECUBE-160731 position.

\begin{figure} [t!]
  \centerline{
    \includegraphics[width=12cm]{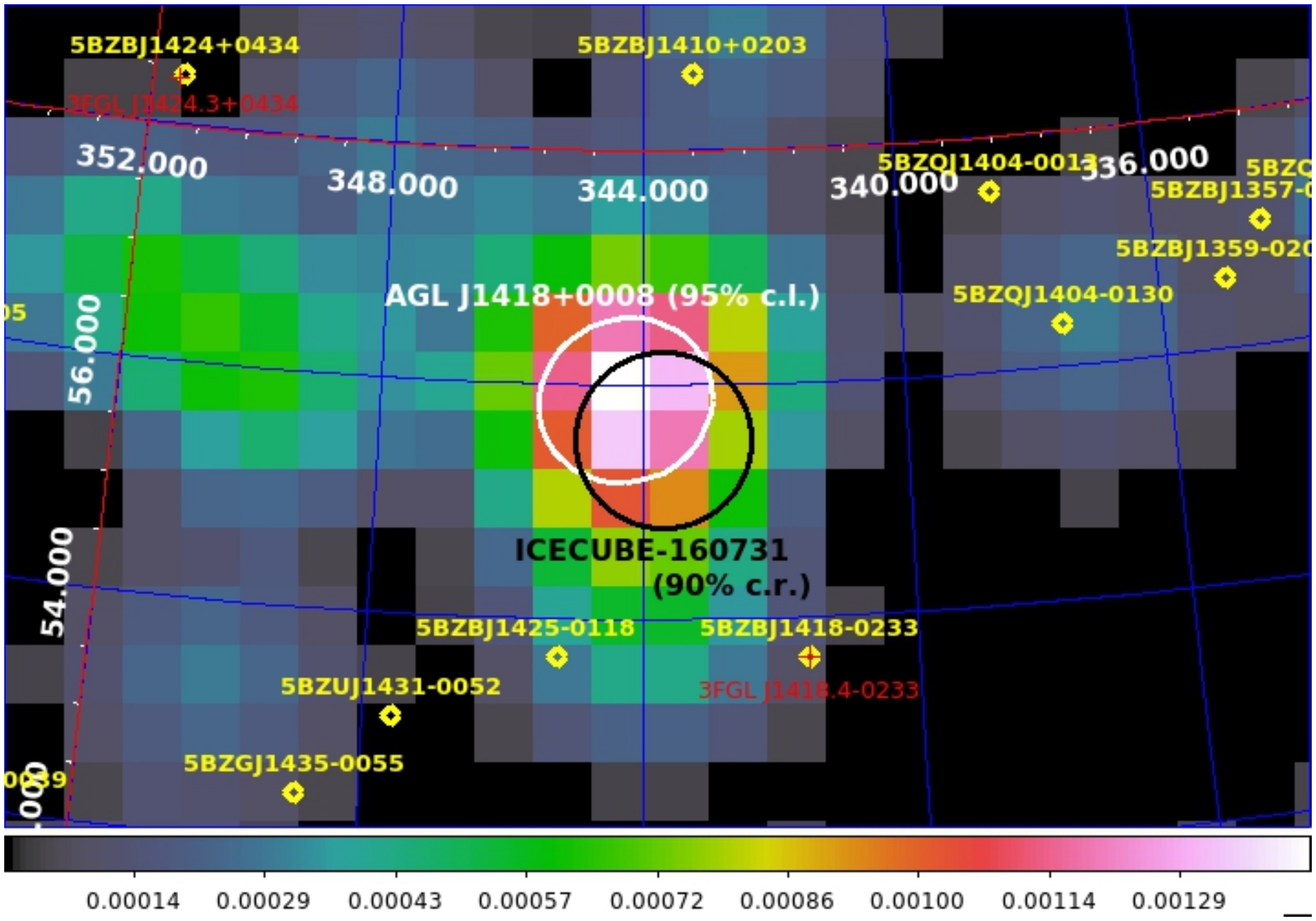}
  }
  \caption{{\agile}-GRID intensity map in [$\mathrm{ph~cm^{-2}~s^{-1}~sr^{-1}}$] 
and Galactic coordinates, centered at the ICECUBE-160731 position, 
from $T_0-1.8$ to $T_0-0.8$~days
The black circle shows the 90\% c.r. of the neutrino event 
while the white circle shows the 95\% C.L. ellipse contour corresponding to the 
{\agile}-GRID ML detection, AGL J1418+0008, described in the text. The 
classified AGNs from the BZCAT Catalog~\citep{2015Ap&SS.357...75M} 
and the FERMI-LAT sources from the 3FGL Catalog~\citep{2015ApJS..218...23A} 
are shown in yellow and in red, respectively. None of these known 
sources appears within the ICECUBE and {\agile} error circles.}
\label{grid_detection}
\end{figure}

{\bfn The {\agile} {\it a-posteriori} refined analysis} on a 24-hours 
basis shows that the excess is particularly short in time, mostly 
concentrated between July 29th and 30th, 2016.
{\bfn By examining the arrival times of the gamma event file, we found a 
a clusterization of five counts in less than 7 hours around ($T_0$-1)~day 
within 1.5 degrees from the ICECUBE centroid}. 
\textcl{\bfn In particular, on the 24-hours integration 
from MJD 57598.25 to 57599.25 ($(T_0-1.8; T_0-0.8)$~days), which fully 
contains the event clusterization,} we obtained a ML significance 
of the peak \gray emission of 4.9$\sigma$ at the Galactic centroid coordinates: 
l,b=(344.26, 55.86) $\pm$ 0.8 [deg] (95\% stat. c.l.) $\pm$ 0.1 [deg] (syst.), 
with a flux F(E$>$100~MeV)=$(3.5 \pm 1.3) \times 10^{-6} \, \rm ph \, cm^{-2} \, s^{-1}$. 

\textcl{\bfn The new {\agile} transient, named AGL J1418+0008, 
positionally consistent with the ICECUBE-160731 error circle might 
then be a possible precursor of the neutrino event.}

Figure~\ref{grid_detection} shows the {\agile}-GRID intensity map centered 
at the ICECUBE-160731 position, in the 24-hours time interval 
correspondent to the peak significance. The white region defines 
the 95\% C.L. ellipse contour of the {\agile}-GRID detection AGL J1418+0008, 
which is well compatible with the ICECUBE-160731 
90\% c.r. error circle (black circle). Figure~\ref{grid_detection} 
shows also the position of the known sources from the 5th edition of 
the BZCAT and FERMI-LAT 3FGL catalogs~\citep{2015Ap&SS.357...75M, 2015ApJS..218...23A}. 
None of these known sources lies within the {\agile} or ICECUBE error circles.
A further search in the Second and Third FERMI-LAT high-energy 
sources Catalogs (2FHL and 3FHL, \cite{2016ApJS..222....5A, 2017arXiv170200664T}) 
does not show again any possible association with known \gray counterparts. 
The closest 3FHL source is 3FHL J1418.4-0233 (associated to the 
BL Lac blazar 5BZB J1418-0233~\citep{2015Ap&SS.357...75M}), which is more 
than 2$^\circ$ away from the neutrino position.

\subsection{Search for gamma-ray emission in {\agile} archival data}
The whole public {\agile}-GRID archival data from Dic. 2007 up to Nov. 2016
have been investigated in order to search for other possible 
previous and later \gray transient episodes around the 
ICECUBE-160731 position.
This long time-scale search has been performed by using 
the {\agile}-LV3 online tool~\citep{AGILELV3} accessible from 
the ASDC Multi-Mission Archive (MMIA) web pages\footnote{URL: 
http://www.asdc.asi.it/mmia/index.php?mission=agilelv3mmia}.
This tool allows fast online interactive analysis based 
on the Level-3 (LV3) {\agile}-GRID archive of pre-computed counts, 
exposure and diffuse background emission maps.

The search for transient emission above 100 MeV on 2-day 
integration times did not show any other significant 
detection but the one compatible with the {\agile} {\it QL} result 
between one and two days before $T_0$ (over a total of 271 analyzed maps).

We finally performed a ML analysis centered on the ICECUBE 
position using the LV3 pre-computed maps for the whole AGILE 
observing time (9 years). We obtained an UL of 
$3.5 \times 10^{-8} \, \rm ph~cm^{-2}~s^{-1}$ 
(E$>$100~MeV, for a 95\% C.L.).

\section{Multi-wavelength follow-up of ICECUBE-160731}
The ICECUBE-160731 detection triggered a thorough campaign of 
MWL follow-up observations. These observations covered a large 
part of the entire e.m. spectrum, from the optical band 
(Global MASTER net, iPTF P48, LCOGT) to the VHE gamma-rays 
(HAWC, MAGIC, HESS, ...).

Very few observatories and space missions were observing 
the neutrino event position to $T_0$. Apart from {\agile} 
and facilities like HAWC, ANTARES and FERMI-LAT, which have 
access to a large part of the sky almost the whole day, all the 
others had to re-point to the ICECUBE position 
a few minutes or even hours after $T_0$. In this section, we 
will summarize the most interesting results of the MWL 
follow-up, reminding the reader to the 
Appendix~\ref{app_mwl_followup} for a summary of 
all other observations published in ATel and 
GCNs in the hours or the days after the event.

\begin{figure} [t!]
  \centerline{
   \includegraphics[width=12cm]{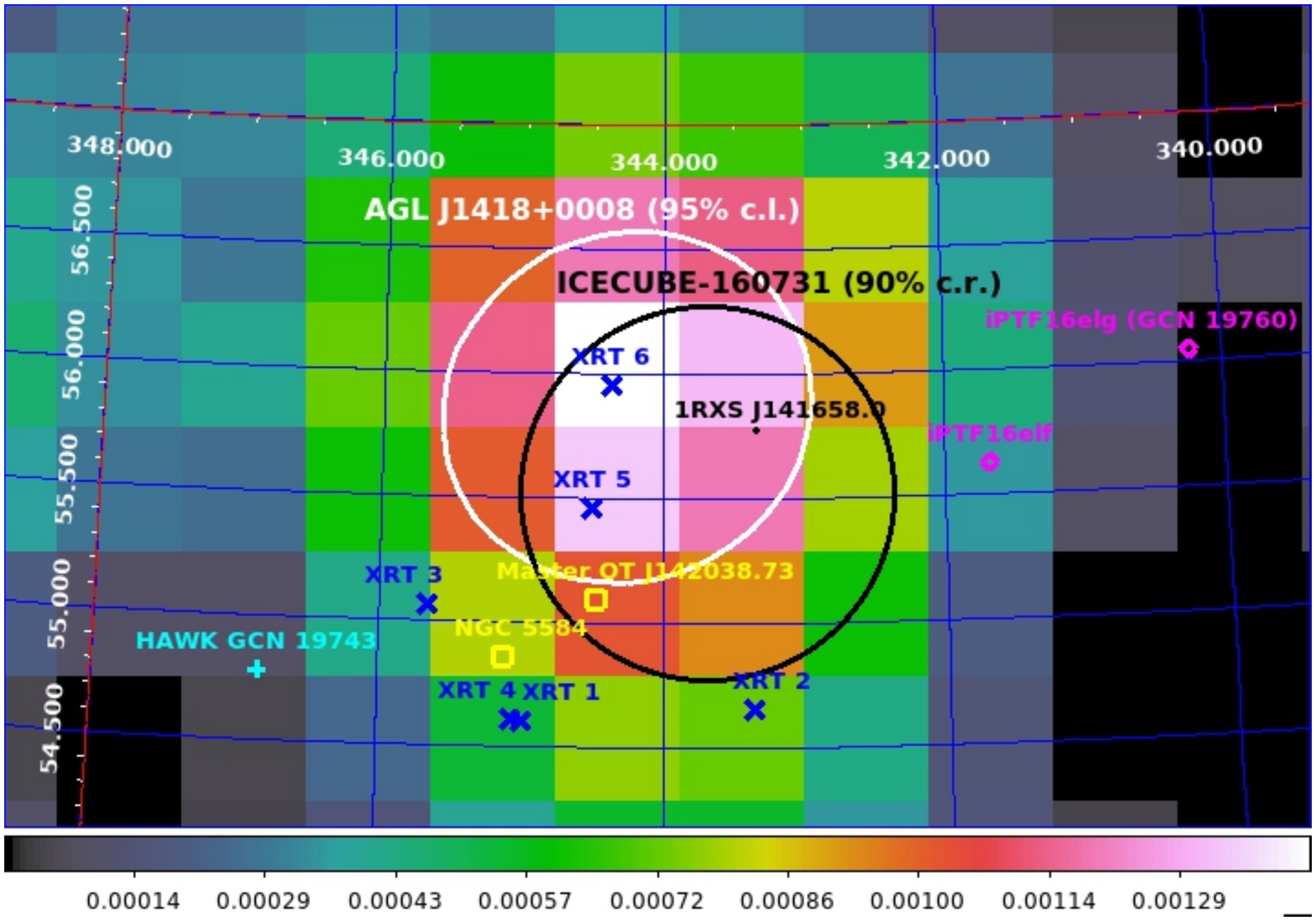}
  }
  \caption{{\agile}-GRID intensity map in [$\rm ph~cm^{-2}~s^{-1}~sr^{-1}$] zoomed 
around the ICECUBE-160731 position, in the time interval $(T_0-1.8; T_0-0.8)$~days. 
The black and white circles again show, respectively, the 90\% c.r. of 
the ICECUBE event and the 95\% C.L. contour of the {\agile}-GRID 
detection AGL J1418+0008. The figure shows also the positions of 
several e.m. candidates found during the MWL follow-up. 
Cyan cross: HAWC best archival search result~\citep{2016GCN..19743...1T}; 
blue crosses: the six SWIFT-XRT sources reported in~\citep{2016GCN..19747...1E, 
2016ATel.9294....1E}; yellow boxes: two optical sources 
(one steady, one transient) detected by the Global MASTER net~\citep{2016ATel.9298....1L, 
2016GCN..19748...1L}); magenta diamonds: two optical transients detected by 
iPTF P48~\citep{2016GCN..19760...1S}. Black point: the X-ray source 1RXS J141658.0-001449, 
which appears within both error circles, is one of the best neutrino-emitter candidate 
found in the additional search made with the ASDC tools described in the text.}
\label{agile_mwl_followup}
\end{figure}

In the X-ray band, SWIFT observed the ICECUBE-160731 error circle 
region starting approximately from $(T_0 + 1)$~hrs till 
$(T_0 + 12)$~hrs~\citep{2016GCN..19747...1E, 2016ATel.9294....1E}. 
The XRT instrument on-board of the SWIFT satellite detected six 
sources in the 0.3-10~keV band. Figure~\ref{agile_mwl_followup} shows a zoom 
of the {\agile}-GRID intensity map over the integration of 
the {\agile} peak detection, with the location of the 
six SWIFT-XRT sources, numbered 1 to 6 (blue crosses in 
Fig.~\ref{agile_mwl_followup}). After the revision of the 
best-fit neutrino arrival direction and its error radius, three 
of the detected XRT sources eventually lay outside
the revised ICECUBE-160731 error circle. Only sources \#5 and 
\#6 are still compatible with the neutrino position 
(and within the {\it AGILE} ellipse contour), while source 
\#2 remains just on the border.

In the optical region, the Global MASTER Optical Network 
performed a search for optical transients in the time interval 
$(T_0 + 17; T_0 + 21)$~hrs~\citep{2016ATel.9298....1L, 2016GCN..19748...1L}.
They only detected a point-like event, classified as 
MASTER OT J142038.73-002500.1, that might have been induced 
by particle crossing the CCD, and the bright NGC 5584 galaxy 
(which, anyhow, is already outside the revised error circle) 
(yellow boxes in Fig.~\ref{agile_mwl_followup}). 
Rapid follow-up observations in the Optical/IR band, started 
only 3.5 hours after $T_0$, were performed 
by the Palomar 48-inch telescope (iPTF P48)~\citep{2016GCN..19760...1S}. 
They detected two optical transient candidates at 1.1 and 2.0$^\circ$ 
from the initial neutrino candidate position (magenta 
diamonds in Fig.~\ref{agile_mwl_followup}).

In the gamma-ray band, FERMI-GBM could not observe the region 
at $T_0$ since the position was occulted by the Earth~\citep{2016GCN..19758...1B} 
while FERMI-LAT reported only flux ULs (95\% C.L.) above 
100 MeV of $10^{-7} \rm ph~cm^{-2}~s^{-1}$ 
in 2.25 days of exposure starting from a 2016-07-31 00:00 UTC, and 
of $0.6 \times 10^{-7} \rm ph~cm^{-2}~s^{-1}$ in 
8.25 days of exposure starting from 2016-07-25 at 
00:00 UTC~\citep{2016ATel.9303....1C}. 
{\bfn As shown in Appendix~\ref{AgilevsFermi}, the non-detection of 
any \gray precursor by Fermi-LAT might be due to a low exposure of 
the ICECUBE region during the {\agile} \gray transient.}

At the time of the neutrino event $T_0$, the INTEGRAL satellite, 
which also has the capability to cover almost the whole 
sky~\citep{2016ApJ...820L..36S}, was not observing because 
it was close to perigee inside the Earth radiation belts.

The ICECUBE region was also observed in the VHE band by several 
experiments (see Appendix~\ref{app_mwl_followup} for details). 
Apart from HAWC, that has a 24-hours duty cycle, all the others 
could re-point to the ICECUBE position hours later than $T_0$, 
reporting only flux ULs above different energy thresholds. On a 
search for steady source using archival data, the HAWC Collaboration  
reported about a location with a pre-trial significance of 
3.57$\sigma$ at R.A.,Dec (J2000)=(216.43, 0.15) [deg]~\citep{2016GCN..19743...1T} 
(shown as cyan cross in Fig.~\ref{agile_mwl_followup}), although more 
than 2$^\circ$ away from the neutrino error circle. {\bfn Considering 
the number of trials quoted in the HAWC GCN, this is not a significant detection.}

\section{Possible neutrino-emitter e.m. sources in the ICECUBE-160731 
and {\agile} AGL J1418+0008 error regions}
\textcl{\bfn In what follows, we will further 
investigate whether some of the steady/transient sources found during 
the MWL follow-up are good candidates as the ICECUBE-160731 emitter. 
In particular, we decided to review only the e.m. sources 
still within the revised ICECUBE error region plus the 
closest optical transient detected by iPTF48 (named iPTF16elf, 
\cite{2016GCN..19760...1S}) (see Fig.~\ref{agile_mwl_followup}). 
Table~\ref{table_x-opt_candidates} shows the main characteristics 
of the five e.m. sources satisfying the chosen selection criteria. 
The table also shows the most likely known association as reported 
from each of the ATel announcing the detection obtained during the follow-up.}

\begin{table}[t!]
\caption{\bf Optical and X-ray sources detected within the revised 
ICECUBE-160731 error circle during the MWL follow-up}
\label{table_x-opt_candidates}
\vskip .3cm
\scriptsize
\begin{minipage}{10cm}
\begin{tabular}{|c|c|c|c|c|c|}
\hline
Mission/Observatory & Source ID/name\footnote{See Fig.~\ref{agile_mwl_followup}.} & R.A. (J2000) & Dec (J2000) & Association & Class \\
 & [deg] & [deg] & & & \\
\hline
SWIFT-XRT (ATel \#9294) & XRT \#2 & 214.90209 & -1.145917 & 2QZ J141936.0-010841 & quasar \\
SWIFT-XRT (ATel \#9294) & XRT \#5 & 214.95898 & -0.11266  & 2QZ J141949.8-000644 & quasar \\
SWIFT-XRT (ATel \#9294) & XRT \#6 & 214.61169 &  0.24144  & 2MASS J14182661+0014283 & star \\
Global MASTER net (ATel \#9298) & OT J142038.73-002500.1\footnote{The astrophysical origin 
of this transient is not confirmed.} & 215.161375 & -0.416694 & SDSS J142041.62-002413.1 & galaxy \\
iPTF P48 (GCN 19760) & iPTF16elf & 213.555124 & -0.894361 & Z 18-88 & galaxy \\
\hline
\end{tabular}
\end{minipage}
\end{table}

\textcl{\bfn To find some of the key features of one 
of the most promising neutrino-emitter candidates, the High-energy 
peaked BL Lac (HBL) types of AGNs~\citep{2016MNRAS.457.3582P, 2017MNRAS.468..597R}, 
we reviewed the initial counterpart association and, moreover, we 
investigated the broad-band spectral properties of each object.}

\textcl{\bfn The first two SWIFT-XRT sources detected during the 
follow-up, \#2 and \#5~\citep{2016GCN..19747...1E}, are consistent 
with the position of two known quasars:} source \#2 is 9.12" from 
2QZ J141936.0-010841\footnote{Also known as [VV2010] J141936.0-010840 
(VV2010 Cat., \cite{2010A&A...518A..10V}) and SDSS J141935.99-010840.2 
(SDSS Cat. -- Release \#7, \cite{2009ApJS..182..543A}).} 
(2QZ Cat, \cite{2001MNRAS.322L..29C}) while source \#5 is 4.5" from 
2QZ J141949.8-000644\footnote{Also known as [VV2010] 
J141949.9-000644~\citep{2010A&A...518A..10V}, 2MASS J14194982-0006432 
(2MASS Cat., \cite{2003tmc..book.....C}), and SDSS J141949.83-000643.7~\citep{2009ApJS..182..543A}.}. 
\textcl{\bfn By looking to their spectral energy distributions (SEDs), 
built using both the XRT detections and MWL archival data, 
neither of the two quasars shows} hints of high-peaked synchrotron emission, which is 
one of the key feature used to identify a HBL type of AGN. Moreover, they 
completely lack radio emission, which leads us to conclude that they might be 
radio-quiet quasars \textcl{\bfn and we can discard them as 
possible emitter of the ICECUBE-160731 neutrino}.

XRT source \#6 is $\sim 2.5"$ from 2MASS J14182661+0014283, a known 
G-type star, and \textcl{\bfn it thus can as well be excluded as 
possible source candidate of the neutrino emission.}

Concerning the two optical transient candidates OT 
J142038.73-002500.1 and iPTF16elf, they are both positionally 
consistent with two galaxies (respectively, SDSS J142041.62-002413.1 
(z=0.054) and Z 18-88 (z=0.038)), which form part of a cluster. For 
both, there are no evident indications of blazar features in their 
respective SEDs.

\begin{figure} [t!]
  \centering
  \includegraphics[trim=10 50 15 15, clip, width=14cm]{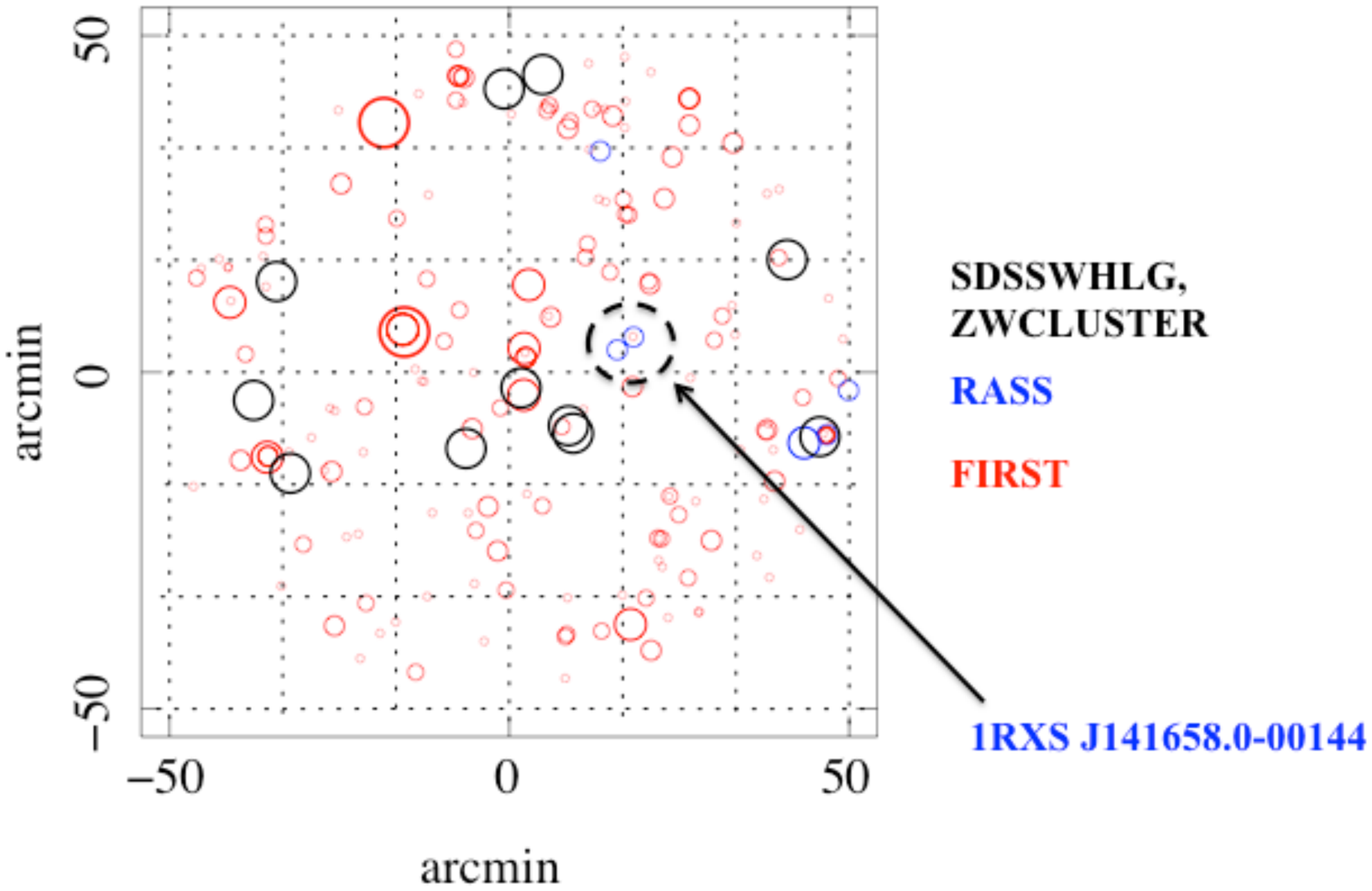}
  \caption{R.A.-Dec sky map (J2000) obtained with the ASDC 
{\it SkyExplorer} tool showing known radio, optical and X-ray sources within 50 arcmin 
from the ICECUBE-160731 position. The map also covers most of the 
95\% C.L. error circle of the {\agile} detection described in Sect.~\ref{precursor}. 
Black circles show sources from the SDSSWHLGC and the ZWCLUSTER 
catalogs~\citep{2009ApJS..183..197W, 1961cgcg.book.....Z}; 
blue circle sources from the ROSAT All Sky Survey (RASS) 
catalogs~\citep{1999A&A...349..389V, 2000IAUC.7432R...1V}; 
red circles are radio sources from the FIRST survey at 
1.4 GHz~\citep{1997ApJ...475..479W}. The dashed 
circle indicates the position of the RASS 1RXS J141658.0-001449 source and 
the nearby FIRST 1.4 GHz radio-source (blue circle with smallest red 
circle inside), a possible HBL AGN candidate (see text for details).}
\label{asdcexplorer_search}
\end{figure}

\textcl{\bfn Besides the review of five e.m. candidates 
found during the ICECUBE-160731 MWL follow-up, we searched for other 
possible counterparts within the ICECUBE 90\% error circle 
by exploring the ASDC resident and external catalogs 
using the online ASDC {\it SkyExplorer} tool\footnote{https://tools.asdc.asi.it}.} 
In particular, we focused our search to known \textcl{\bfn radio and X-ray sources} 
which might show the typical characteristics of HBL/HSP AGN 
blazars~\citep{2017A&A...598A..17C}: low radio fluxes and low IR-radio 
spectrum slopes; {\bfn high X-ray-to-radio flux ratios;} $\nu$ synchrotron peaks above $10^{15}$~Hz.
 
A query of 50 arcmin around the ICECUBE-160731 centroid Galactic coordinates 
l,b=(343.68, 55.52 deg) selecting, among others, radio and X-ray sources from 
the FIRST~\citep{1997ApJ...475..479W} and the RASS 
Catalogs~\citep{1999A&A...349..389V, 2000IAUC.7432R...1V}, 
returns several objects (see Figure~\ref{asdcexplorer_search}). 
\textcl{\bfn Following the search criteria defined above, one 
of the most interesting object} resulting from the query is a RASS source 
appearing at $\sim$19 arcmin from the center, 1RXS J141658.0-001449, 
\textcl{\bfn with position and related uncertainty R.A.,Dec (J2000)=($\rm 14^{h}16^{m}58^{s}.0, 
-00^\circ14'49"$) $\pm$ 25", (indicated by the dashed circle in 
Fig.~\ref{asdcexplorer_search}). This cataloged 
X-ray source is the only one in the field showing a FIRST {\bfn weak} 
radio source (F=1.99 mJy; R.A.,Dec (J2000)=($\rm 14^{h}16^{m}58^{s}.27, 
-00^\circ14'44.87"$)) within its error circle.} 
\textcl{\bfn A further search in the ASDC optical catalogs found} 
a faint galaxy, SDSS J141658.90-001442.5 (mv $\sim$23), at 9.6 arcsec 
from the FIRST source (14.8 arcsec from the RASS source).

\begin{figure} [t!]
  \centering
  \includegraphics[width=12cm]{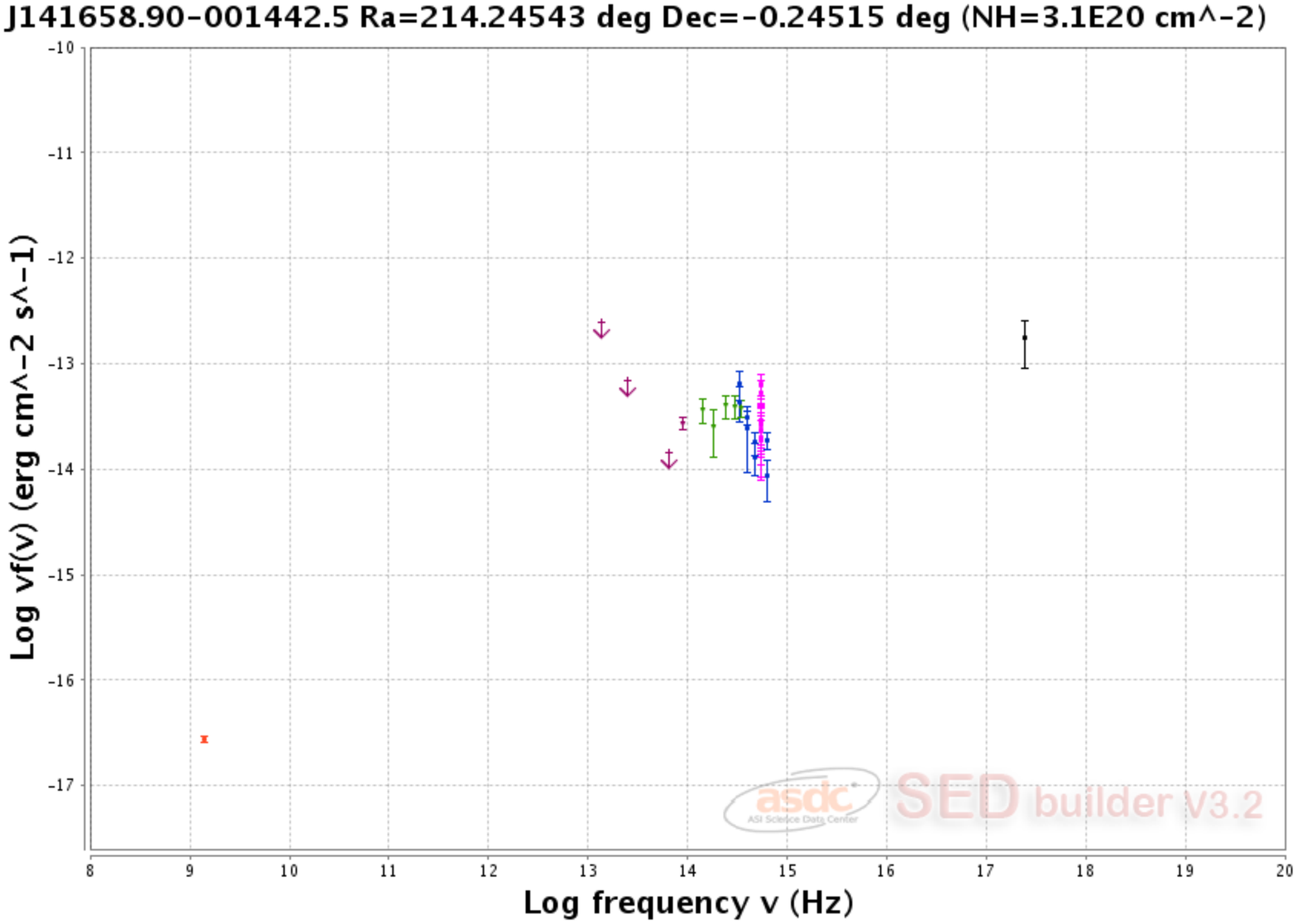}
\caption{Spectral energy distribution (SED) of the possible HBL candidate, 
the faint SDSS J141658.90-001442.5 galaxy, found within the ICECUBE-160731 
error circle. The galaxy appears within the 25" error circle of the 
RASS source 1RXS J141658.0-001449 ($\nu~F_{\nu}$ value shown as 
black point in the SED), along with a FIRST 2~mJy radio source (red point). 
Optical and IR data of the SDSS J141658.90 galaxy are from: 
Sloan Digital Sky Survey (SDSS) -- Release \#7 and \#13 
(blue points, \cite{2009ApJS..182..543A, 2016arXiv160802013S}); 
Catalina Real-Time Transient Survey (CRTS) (magenta points, \cite{2009ApJ...696..870D}); 
VIKING survey (green points, \cite{2013Msngr.154...32E}); 
AllWISE Data Release (purple points, \cite{2014yCat.2328....0C}).}
\label{SED_giommicandidate}
\end{figure}

Assuming the radio/optical/X-ray emission comes from the same galaxy, we have 
produced the SED shown in Figure~\ref{SED_giommicandidate}. 
{\bfn The high value of the ratio between the 1RXS J141658.0-001449 
flux density in the 0.1-2.4 keV band and the FIRST radio source $\nu~F_{\nu}$ value 
at 1.4~GHz (respectively, black and red points in Fig.~\ref{SED_giommicandidate}) 
might hint to a non-thermal synchrotron emission peaking above $10^{15}$~Hz, typical of 
a HBL AGN blazar.} Considering these types of e.m. sources as the most 
likely neutrino-emitters, the X-ray source 1RXS J141658.0-001449 
(and the plausible host galaxy SDSS J141658.90-001442.5) appears as one 
of the candidate as origin of the ICECUBE-160731 event. 

\textcl{\bfn This source was not in the field covered by the July 31st, 2016, SWIFT 
series of ToO observations~\citep{2016GCN..19747...1E}.} Interestingly, the 
source lies also within the 95\% error ellipse contour of the {\it AGILE} detection 
occurred before the neutrino event time $T_0$ (see Fig.~\ref{agile_mwl_followup}).

\subsection{SWIFT ToO data on the 1RXS J141658.0-001449 field}
\begin{figure} [t!]
  \centerline{
  \includegraphics[width=10cm]{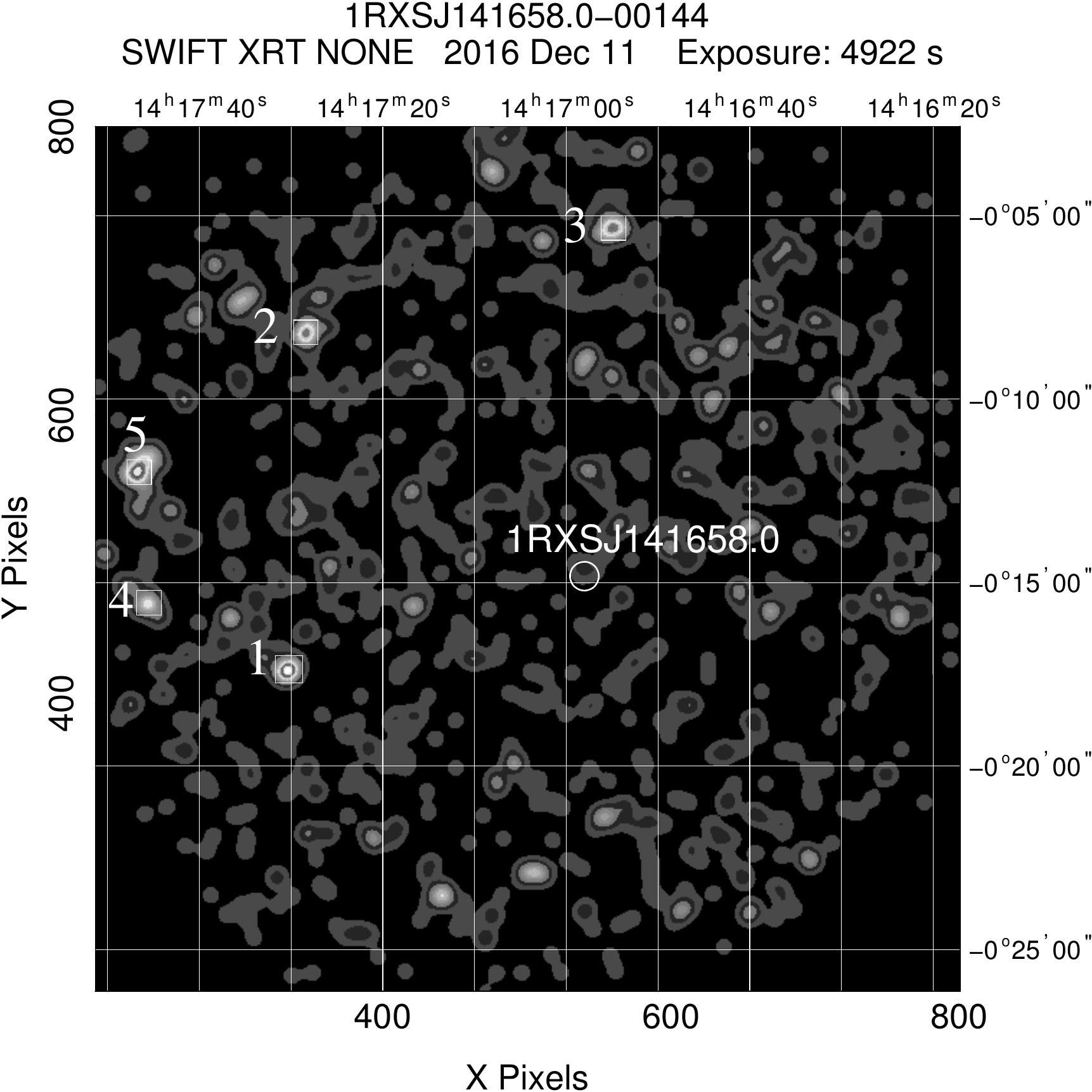}
}
\caption{Smoothed SWIFT-XRT count map (0.3--10 keV) centered on the 
ROSAT/RASS-FSC 1RXS J141658.0-001449 source, obtained from 
the SWIFT ToO executed on Dic. 2016, almost six months later than 
the ICECUBE-160731 neutrino detection. Total exposure: $\sim4.9$~ks. 
White boxes show the 5 field sources detect by using the XIMAGE {\it detect} 
algorithm. No significant X-ray excess is found at the 1RXS J141658.0 position.} 
\label{XRT_ctsmap}
\end{figure}

To better estimate the position and the spectrum of the RASS 
1RXS J141658.0-001449 source (which was not in the field covered by 
the first SWIFT series of ToO observations~\citep{2016GCN..19747...1E}) 
and determine a stronger spatial correlation with the radio and 
optical sources described above, a new SWIFT ToO has been submitted 
and executed in December 2016, almost six months later than 
the ICECUBE-160731 neutrino detection.

The data were collected in five distinct $\sim1$~ks exposures centered on the 
1RXS J141658.0 source position between 2016-12-11 00:32:59 UT
and 2016-12-15 07:07:53 UT and are entirely in Photon Counting (PC) mode\footnote{
Correspondent SWIFT OBSERVATION IDs: from 00034815001 to 00034815005.}.

Figure~\ref{XRT_ctsmap} shows the (smoothed) cumulative XRT count map in the 0.3-10 keV 
energy range, with an overall exposure of 4.9~ks. The position 
of the 1RXS J141658.0 source (with its quoted error circle) is 
superimposed to the map (white circle near the map center). No apparent 
X-ray excess is visible at the 1RXS J141658.0 position.

Using the XIMAGE {\it sosta} algorithm, we derive a 3$\sigma$ UL of 
$3.1 \times 10^{-3} \mathrm{cts~s^{-1}}$ in the XRT energy 
band on the 1RXS J141658.0 position. Assuming a source with a 
power-law photon index of 1.7, we evaluated an upper limit of 
$4.6 \times 10^{-3} \mathrm{cts~s^{-1}}$ in the ROSAT PSPC band. 
This value is well below the count rate of $(2.19 \pm 1.04) \times 10^{-2}$ 
quoted for 1RXS J141658.0-001449 in the RASS-FSC Catalog. 
This might indicate an intrinsic variability of the source, 
which was significant only during the RASS observation. 
It should be noted that this source does not appear anymore in the second 
ROSAT all-sky survey (2RXS) Catalog~\citep{2016A&A...588A.103B}, an 
extended and revised version of the 1RXS Catalog that 
contains a significant reduced number of low reliability sources.

Applying the XIMAGE {\it detect} algorithm on the overall 5~ks XRT count map, 
weighted by the correspondent sum of each single XRT exposure, five 
(uncataloged) X-ray field sources are detected within the FoV (see Fig.~\ref{XRT_ctsmap}). 
Table~\ref{table_xrtToOsources} reports count rates, source coordinates, SNR ratio and 
probability to be a background fluctuation for all the five detections. 
Studies of the characteristics of the five field sources is ongoing.

\begin{table}[t]
\begin{center}
\caption{\bf SWIFT-XRT detections in the 0.3--10 keV band from the ToO centered 
on the 1RXS J141658.0-001449 source}
\label{table_xrtToOsources}
\vskip .3cm
\begin{tabular}{|r|c|c|c|c|c|}
  \hline
ID & Count rate & R.A. (J2000) & Dec (J2000) & prob. & SNR \\
\, & [$\rm cts~s^{-1}$] & [hh mm ss] & [dd mm ss] & & \\
\hline
1 & 3.16E-03$\pm$1.1E-03 & 14 17 30.209 & -00 17 21.842 & 6.541E-08 & 2.9 \\
2 & 2.33E-03$\pm$8.8E-04 & 14 17 28.391 & -00 08 10.772 & 1.884E-06 & 2.7 \\
3 & 3.06E-03$\pm$1.1E-03 & 14 16 54.849 & -00 05 20.036 & 2.425E-07 & 2.8 \\
4 & 2.75E-03$\pm$1.1E-03 & 14 17 45.479 & -00 15 32.932 & 5.285E-06 & 2.5 \\
5 & 4.39E-03$\pm$1.4E-03 & 14 17 46.553 & -00 11 59.302 & 2.972E-10 & 3.2 \\
\hline
\end{tabular}
\end{center}
\end{table}

\section{Discussion and conclusions}
We reported the results of {\agile} \gray observations of the ICECUBE-160731 
neutrino event error region. These observations covered the event sky 
location at the event time $T_0$ and also allowed us to search for 
e.m. \gray counterparts before and after the event.

The analysis of the {\agile}-GRID data in the time window 
$T_0 \pm 1$~ks {\bfn with the {\agile} burst search system}
has not shown any significant gamma-ray excess above 
30~MeV from the neutrino position. Moreover, no burst-like 
events using the {\agile}-MCAL and the AC ratemeters around $T_0$ 
have been detected. 
{\bfn Instead, an automatic detection above 100~MeV, 
compatible with the ICECUBE position, appeared from 
the {\agile} {\it QL} procedure on a predefined 
48-hours interval centered around one day and a half before $T_0$. 
Considering all the number of trials performed by the {\agile} {\it QL} 
system and the chance probability to have a \gray excess in coincidence 
with the neutrino position, the automatic detection reaches a 
combined post-trial significance of about 4$\sigma$. 
A refined data analysis confirms the {\it QL} detection already 
reported in the ATel \#9295~\citep{2016ATel.9295....1L}. 
This new {\agile}-GRID \gray transient, named AGL J1418+0008, is 
rather concentrated in time, showing a clusterization of 
events around ($T_0$-1)~days, and reaching a peak ML significance of 
4.9$\sigma$ on the 24-hours integration covering the 
interval ($T_0$-1.8; $T_0$-0.8)~days. AGL J1418+0008 thus stands as 
possible ICECUBE-160731 gamma-ray precursor.}

No other space missions or observatories have reported any 
clear indication of a transient e.m. emission consistent 
with the neutrino position and time $T_0$. This non-detection 
of an e.m. counterpart in any of the wavelengths covered by the ICECUBE-160731 
follow-up does not exclude the possibility of a bright rapid gamma-ray 
flare precursor just before the neutrino detection. 
Most of the instruments involved in the e.m. follow-up, in fact, 
could re-point their instruments only hours or even a day after $T_0$, 
and might have missed the flaring episode seen by {\agile} at E$>$100~MeV.

As said in the MLW follow-up summary, FERMI-LAT did not report any evidence 
of a precursor above 100 MeV. As we show in Appendix~\ref{AgilevsFermi}, this 
might be due to a very high FERMI-LAT observing angle and a very low exposure 
of the ICECUBE region with respect to the {\agile} observations.

Given the high Galactic latitude of the ICECUBE neutrino arrival 
direction (b=55.52 [deg]), we do expect an extra-galactic origin of this 
event. Indeed, several authors (i.e., \cite{2014PhRvD..90d3005A, 2016MNRAS.457.3582P}) 
assume that blazar AGNs are the main VHE neutrino-emitter candidates 
and the only sources able to explain the common origin of 
the diffuse neutrino background seen by ICECUBE, the extra-galactic 
cosmic-ray component and the isotropic diffuse gamma-ray background 
observed by FERMI~\citep{2015ApJ...799...86A}. \cite{2016NatPh..12..807K} 
found for the first time a significant probability that one of 
the ICECUBE PeV event was spatially and temporally coincident 
with a major gamma-ray outburst of the Flat Spectrum Radio Quasar (FSRQ) 
PKS B1424-418. Considering that there is a substantial fraction of the blazar 
population not resolved yet, Kadler et al. estimate 
that around 30\% of the detected multi-TeV/PeV neutrinos will not be 
associated with any known \gray blazar, like appears to be
the case of the ICECUBE-160731 event.

Recently, ~\cite{2017MNRAS.468..597R} found that a significant 
correlation between known HBL blazars, ICECUBE neutrinos and 
UHECRs detected by Auger and the Telescope Array (TA) exists. 
We thus searched for a HBL candidate counterpart inside the common ICECUBE and 
{\agile} AGL J1418+0008 error circles and found a possible HBL source, 
the Sloan faint galaxy SDSS J141658.90-001442.5, which appears 
within the positional error of the RASS source 1RXS J141658.0-001449 and close 
to a FIRST 2~mJy radio source. The ICECUBE-160731 SWIFT follow-up, 
although rapid, did not cover the field around this possible 
e.m. candidate. A new SWIFT ToO then has been submitted in order 
to characterize better this RASS-FSC source. Unfortunately, the 
ToO was performed about 6 months after the neutrino event, and 
the analysis of the XRT data from the almost 5~ks exposure 
did not reveal any significant X-ray emission at the 1RXS J141658.0 position, 
providing a 3$\sigma$ UL of $3.1 \times 10^{-3}~\mathrm{cts~s^{-1}}$ in the 
0.3--10 keV band. We then cannot confirm at the moment our hypothesis 
about the HBL nature of this source that, anyhow, might 
have been detected during the ROSAT survey because in an intrinsic 
X-ray high-state.

Other possible PeV neutrino-emitters have been proposed, like Starburst galaxies, 
giant radio galaxies with misaligned jets, gamma-ray bursts (GRBs) 
(see \cite{2014PhRvD..90d3005A} for a review). 
\cite{2016GCN..19888...1L}, for example, 
correlate another recent ICECUBE HESE neutrino event (ICECUBE-160814) 
with an optical transient occurred almost ten days 
after the event time. They postulate the possibility that the neutrino
emitter might be an ejecting white dwarf in a binary system. This is 
an intriguing possibility, although the power budget available in these systems 
(optical companion plus compact object) could not be sufficient to 
accelerate protons up to multi-PeV energies in order to produce sub-PeV/PeV 
neutrinos from $pp$ collisions.

Eventually, none of the other e.m. sources proposed up to now as 
neutrino-emitter candidates are able to explain the bulk of 
multi-wavelength/multi-messenger (neutrinos plus cosmic rays) 
observational data like the HBL/HSP class of blazars~\citep{2017MNRAS.468..597R}.
Indeed, the probability to find a blazar of this class in 
a 1$^\circ$ radius sky-area like the ICECUBE-160731 error 
circle is quite low. Assuming, in fact, an HSP 
density of the order of $5 \times 10^{-2}~\rm \,deg^{-2}$ from the 2WHSP 
catalog~\citep{2017A&A...598A..17C}, there are approximately 
5 HSP/HBL AGNs every 100 squared degrees of sky. Thus, the 
probability to find one of these objects within the roughly 3 
squared degrees covered by the ICECUBE error circle is of about 0.15\%. 
In the specific case of the ICECUBE-160731 neutrino, for example, 
we have not found yet any other potential HBL candidate 
but the one not confirmed with the dedicated SWIFT ToO observations. 
Moreover, the {\agile} transient, not confirmed by FERMI
(although caused by a poor FERMI-LAT visibility just before $T_0$) 
might indicate a possible soft \gray source, in disagreement 
with the hard-spectrum \gray features expected for the HBLs. 

Nevertheless, the HBL scenario can still hold if we assume 
a lepto-hadronic process occurring within the blazar 
jet~\citep{2017A&A...598A..36R}, where the bulk of 
broad-band e.m. emission is due to synchrotron and Inverse Compton 
leptonic processes, while protons would be mainly 
responsible for the neutrino flux (from the decay of 
charged pions produced by photo-meson production on the soft 
photons field within the jet). In this case, \cite{2017A&A...598A..36R} 
foresee that a soft \gray component, peaking at MeV/GeV energies, would 
be expected from re-processing of VHE photons from the decay of $\pi^0$'s 
originated in the $p\gamma$ collisions within the jet. The {\agile} 
observation of the \gray transient AGL J1418+0008, compatible with 
the neutrino position and very close in time to the event $T_0$, 
{\it if} associated with the ICECUBE event, could be then 
explained by such hadronic mechanism.

To conclude, there is also the possibility that the source of the ICECUBE-160731 
neutrino event might be either a different AGN type or a different 
class of source, even though we cannot exclude at the moment a 
moderately bright HBL not yet identified.

\acknowledgments
We would like to thank Paolo Giommi and Matteo Perri, for many fruitful discussions 
and the valuable help with the analysis of the SWIFT ToO, and Paolo Lipari 
for the very useful comments about the paper. We also thank the Swift Team for making 
the SWIFT ToO observations possible, in particular M. H. Siegel 
as the Swift Observatory Duty Scientist. Lastly, we thank the anonymous 
referees for their valuable comments that helped to improve our paper.
AGILE is an ASI space mission with programmatic support from INAF and INFN. We 
acknowledge partial support through the ASI grant no. I/028/12/0. Part of this 
work is based on archival data, software or online services provided by 
the ASI SCIENCE DATA CENTER (ASDC). It is also based on data and/or software 
provided by the High Energy Astrophysics Science Archive Research 
Center (HEASARC), which is a service of the Astrophysics Science Division 
at NASA/GSFC and the High Energy Astrophysics Division of the 
Smithsonian Astrophysical Observatory. This research has also made use 
of the SIMBAD database and the VizieR catalog access tool, operated at 
CDS, Strasbourg, France.

\software{{\agile} scientific analysis software (BUILD 21; \cite{AGILESW}), XIMAGE}

\newpage
\appendix
\section{Comparison between {\it AGILE} and FERMI-LAT data during 
the ICECUBE-160731 event}\label{AgilevsFermi}
\begin{figure} [t!]
  \centering
  \includegraphics[width=12cm]{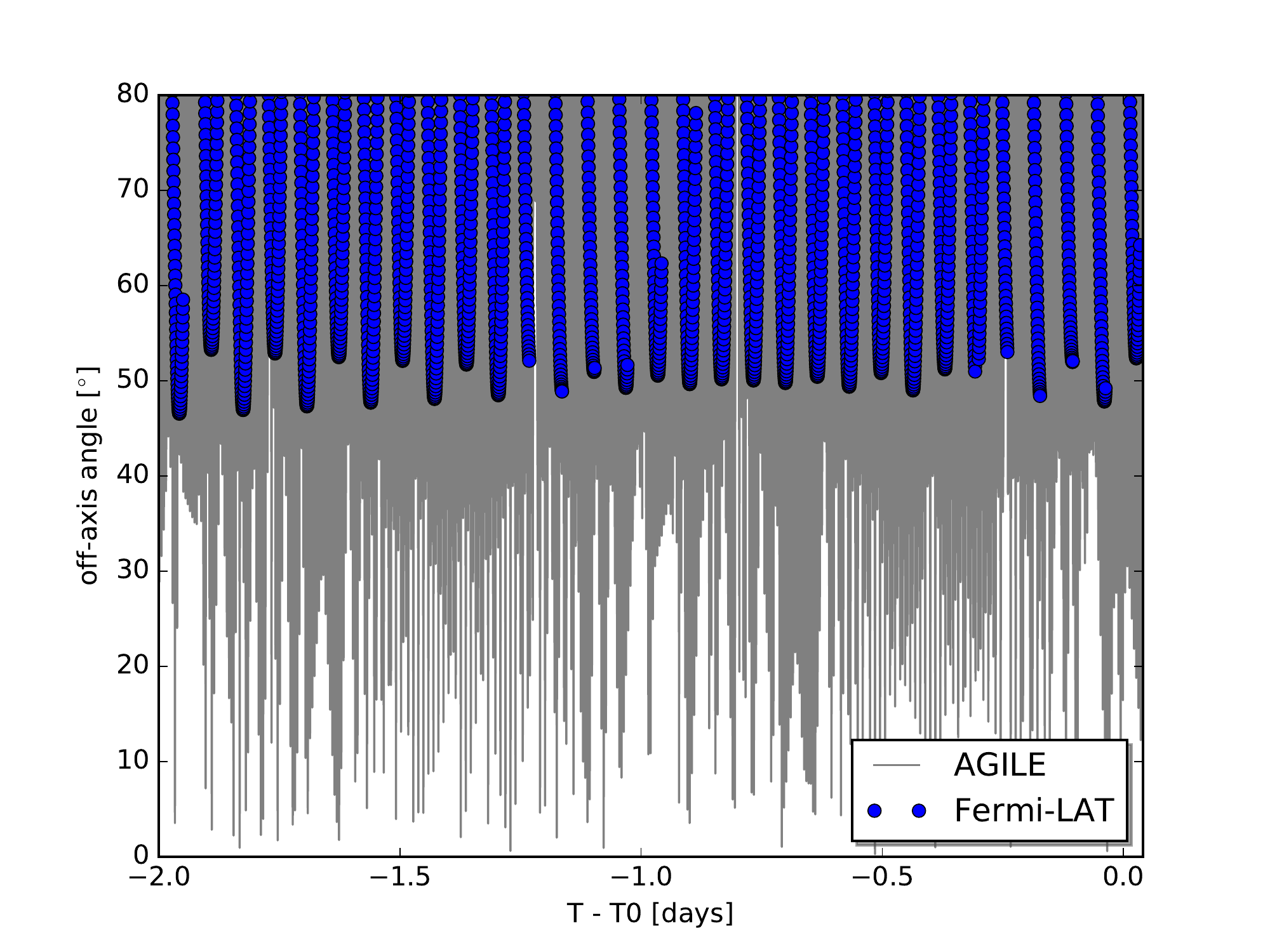}
  \caption{Time-evolution of the ICECUBE-160731 region off-axis angles as observed by 
{\agile} and Fermi-LAT during the 48-hrs time interval ($T_0-2; T_0$)~days (MJD 
57598.07991$\div$57600.07991).}
\label{figure_agile-fermi_vis}
\end{figure}

In this Appendix, we verify that the FERMI-LAT non-detection of the 
{\agile} possible \gray precursor of the neutrino 160731 event might be due 
to a poor exposure and non optimal viewing angle of the ICECUBE 
error circle.

We have compared the FERMI-LAT attitude data with the {\agile} ones 
during the time interval ($T_0-2;T_0$)~days (MJD 57598.07991$\div$57600.07991) 
and found that FERMI-LAT observed the ICECUBE error circle at an off-axis angle 
lower than 50$^\circ$ only for a 3.9\% of its total exposure time, while for {\agile} 
the exposure time below the same off-axis angle amounted to 27.4\% of the total (see 
Figure~\ref{figure_agile-fermi_vis})\footnote{At high values of 
the off-axis angle ($>50^\circ$), the Fermi/LAT sensitivity is up to 50\% lower 
than the nominal on-axis value.}.

Further investigations of the FERMI spacecraft data show also several 
periods of not data-taking during the same time-interval  
(amounting to $\sim15$\% of the total observation time), 
particularly near ($T_0$-1)~days (as it is possible to see 
from Fig.~\ref{figure_agile-fermi_vis}), where {\it AGILE}
found a clusterization of gamma-like events compatible with the ICECUBE error circle.

To prove that during this period the {\agile} and FERMI-LAT exposures 
on the ICECUBE region were {\it at least} comparable, we have evaluated 
the exposures for both instruments on time intervals of 24, 12 and 6 hours 
centered at $(T_0-1)$~days (MJD=57599.07991), where the {\agile} detection 
reached its peak significance.

We downloaded Pass8 data\footnote{From the FERMI data ASDC mirror 
(https://tools.asdc.asi.it).} around 
the position of ICECUBE-160731 and, using version v10r0p5 of the Fermi Science Tools 
provided by the Fermi satellite team\footnote{http://fermi.gsfc.nasa.gov}
and the instrument response function P8R2\_SOURCE\_V6, 
we calculated the mean exposure values on the neutrino error circle 
on those different integration times. We selected Pass8 FRONT and BACK source class 
events and, in order to be comparable with the {\agile} spectral sensitivity 
(optimized to the observation of soft \gray sources with typical 
spectral indexes of 2$\div$2.1), we limited the event energies 
between 0.1 and 10 GeV.

Table~\ref{table_fermi_exps} shows the values of the FERMI-LAT and {\agile} exposures 
on the different time intervals chosen and for a maximum off-axis 
angle between source and FoV center of 50$^\circ$.

\begin{table}[t!]
\begin{center}
\caption{\bf {\agile} and FERMI-LAT exposures on the ICECUBE-160731 error circle
during the period of the detection of the possible \gray 
precursor AGL J1418+0008. For both instruments, a maximum off-axis angle 
of 50$^\circ$ between source and FoV center has been assumed.}
\label{table_fermi_exps}
\vskip .3cm
\begin{tabular}{|c|c|c|c|}
  \hline
Interval & duration & {\agile} mean exp & FERMI-LAT mean exp \\
 $\rm [MJD]$ & [hrs] & ($\rm cm^2 \, s$) & ($\rm cm^2 \, s$) \\
\hline
57598.25 $\div$ 57599.25 & 24 & 3.7E+06 & 3.8E+06 \\
57598.75 $\div$ 57599.25 & 12 & 1.7E+06 & 1.2E+06 \\
57598.875 $\div$ 57599.125 & 6 & 8.2E+05 & 4.7E+05\\
\hline
\end{tabular}
\end{center}
\end{table}

The LAT exposure on the 24-hours interval MJD 57598.25 $\div$ 57599.25 
becomes comparable with the {\agile} exposure of $3.7 \times 10^{6} \rm \, cm^2 \, s$ 
obtained under the same maximum viewing angle and the 
same integration time. On the shorter intervals of 12 and 6 hours 
around ($T_0$-1)~days, the {\agile} exposure becomes even larger 
than the FERMI one. Assuming, thus, a very short \gray flare, as 
the {\agile} detection indicates, it might imply the possibility that 
FERMI, given the very low exposure and the large viewing angle 
of the ICECUBE-160731 position during this period, lost most of 
the \gray transient episode. Differences in the event classification 
algorithms between the two instruments can also bring to a 
detection/non-detection in such cases of short \gray transients 
at the level of $4\sigma$ above the background.

\section{Summary of the ICECUBE-160731 MWL follow-up}\label{app_mwl_followup}

\begin{longrotatetable}
\begin{deluxetable*}{|l|c|c|c|p{2in}|}
\tablecaption{Summary of the MWL follow-up of the ICECUBE-160731 event\label{mwl_followup}}
\tablewidth{700pt}
\tabletypesize{\scriptsize}
\tablehead{
\colhead{Mission/Observatory} & \colhead{ATel} & 
\colhead{GCN Circular} & \colhead{Observation/integration time} & 
\colhead{Comments} \\ 
\colhead{(Energy band)} & \colhead{\#} & \colhead{\#} & \colhead{[UTC]} & \colhead{}
}
\startdata
   HAWC (TeV gamma-rays) & - & 19743 & 2016-07-30 21:28:57 -- 2016-07-31 02:59:15 & No detection around neutrino event time $T_0$ (most significant location (1.12$\sigma$) at R.A., Dec (J2000) = 214.67, 1.04 deg). From archival data, a pre-trial 3.57$\sigma$ detection from R.A., Dec (J2000) = 216.43, 0.15 deg is reported.\\\hline
   SWIFT (X-ray, Optical/UV) & 9294 & 19747 & 2016-07-31 03:00:46 -- 2016-07-31 14:51:52 & Six known or cataloged X-ray sources detected (0.3-10 keV) but no transient events. No transient sources detected in the simultaneous UVOT data.\\\hline
   AGILE (Gamma-rays) & 9295 & - & \begin{tabular}{@{}c@{}}2016-07-29 02:00 -- 2016-07-31 02:00 \\ 2016-07-28 08:00 -- 2016-07-30 08:00\end{tabular} & $>4\sigma$ pre-trials detection on the interval 2016-07-28/2016-07-30 (08:00) UT.\\\hline
   Global MASTER net (Optical) & 9298 & 19748 & From 2016-07-31 19:23:17 on & No optical transients detected inside 2 square degrees around center of ICECUBE-160731 Rev. \#0 error circle. Detected one likely particle CCD event (OT J142038.73-002500.1) and the NGC 5584 galaxy.\\\hline
   FACT (TeV gamma-rays) & - & 19752 & 2016-07-31 21:42 -- 2016-07-31 22:25 & No detection.\\\hline
   HESS (TeV gamma-rays) & 9301 & - & \begin{tabular}{@{}c@{}}2016-07-31/08-01 (1 hr) \\ 2016-08-01/02 (1 hr)\end{tabular} & No detection.\\\hline
   FERMI-LAT (Gamma-rays) & 9303 & - & \begin{tabular}{@{}c@{}}2.25 days from 2016-07-31 00:00 \\ 8.25 days from 2016-07-25 00:00\end{tabular} & No detection above 100 MeV.\\\hline
   FERMI-GBM (X-ray/Gamma-rays) & - & 19758 & Neutrino event trigger time ($T_0$). & Position occulted by Earth at $T_0$. Flux U.L. at 3$\sigma$ level (12-100 keV) on the interval July 30th-Aug. 1st.\\\hline
   iPTF P48 (Optical/IR) & - & 19760 & From 2016-07-31 05:22 on. & No optical transients detected close to the ICECUBE updated error circle. Two optical transients candidate (iPTF16elf and iPTF16elg) detected at 1.1 and 2.0 deg from the neutrino candidate position, both consistent with known galaxies.\\\hline
   MAXI/GSC (X-ray) & 9313 & - &  \begin{tabular}{@{}c@{}}At 2016-07-31 02:32. \\ From 2016-07-20 to 2016-08-03.\end{tabular} & No detection on the 2-20 keV band within the ICECUBE error circle, neither near $T_0$ nor in the period July 20th -- Aug. 3rd. 3$\sigma$ U.L. are provided.\\\hline
   MAGIC (TeV gamma-rays) & 9315 & - & 2016-07-31 21:25 -- 2016-07-31 22:47 & No detection above 600 GeV. \\\hline
   ANTARES (TeV/PeV neutrinos) & 9324 & 19772 & \begin{tabular}{@{}c@{}}$T_0 \pm$~1 hr \\ $T_0 \pm$~1 day\end{tabular} & No up-going muon neutrino candidate events recorded within three degrees of the ICECUBE event coordinates. 90\% U.L. on the fluence from a point-like source are reported.\\\hline
   Konus-Wind (X-ray/Gamma-rays) & - & 19777 & \begin{tabular}{@{}c@{}} $T_0 \pm 1000$s \\ From 5 days before to 1 day after $T_0$.\end{tabular} &  No triggered events detected. 90\% C.L. upper limits are reported on the 20-1200 keV fluence for typical short and long GRB spectra.\\\hline
   LCOGT (Optical) & 9327 & - & From 2016-07-31 23:04:41 till 2016-08-03 18:29:11. & No detection of new optical sources down to 3$\sigma$ limiting magnitudes $>$19.\\
\enddata
\end{deluxetable*}
\end{longrotatetable}

\clearpage
\bibliographystyle{aasjournal}
\bibliography{Lucarelli_AGILE-ICECUBE160731_astroph}

\end{document}